\documentclass[draft]{agujournal2019}
\usepackage{url} 
\usepackage{soul}
\draftfalse

\journalname{JGR: Space Physics}

\begin{document}

%
%


\title{Particle orbits at the magnetopause: Kelvin-Helmholtz induced trapping}

%
%




\authors{M.H.J. Leroy\affil{1}, B. Ripperda\affil{2},
 and R. Keppens\affil{1,3,4}}


\affiliation{1}{Centre for mathematical Plasma-Astrophysics, Department of Mathematics, KU Leuven, Celestijnenlaan 200B, B-3001 Leuven, Belgium}
\affiliation{2}{Institut f\"{u}r Theoretische Physik, Max-von-Laue-Str. 1, D-60438 Frankfurt, Germany}
\affiliation{3}{School of Astronomy and Space Science, Nanjing University, PR China}
\affiliation{4}{Purple Mountain Observatory, Chinese Academy of Sciences, Nanjing, PR China}





\correspondingauthor{Rony Keppens}{rony.keppens@kuleuven.be}




\begin{keypoints}
\item We study charged particle orbits at rolled-up flanks of the magnetopause
\item Kelvin-Helmholtz induced plasma variations cause intricate trapping sites
\end{keypoints}

%
%

%
%


\begin{abstract}
The Kelvin-Helmholtz instability (KHI) is a known mechanism for penetration of solar wind matter into the 
magnetosphere. Using three-dimensional, resistive magnetohydrodynamic simulations, the double mid-latitude reconnection (DMLR) process was shown to efficiently exchange solar wind matter into the magnetosphere, through mixing and reconnection. Here, we compute test particle orbits through DMLR configurations. In the instantaneous electromagnetic fields, charged particle trajectories are integrated using the guiding centre approximation. The mechanisms involved in the electron particle orbits and their kinetic energy evolutions are studied in detail, to identify specific signatures of the DMLR through particle characteristics. 
The charged particle orbits are influenced mainly by magnetic curvature drifts. 
We identify complex, temporarily trapped, trajectories where the combined electric field and (reconnected) magnetic field variations realize local cavities where particles gain energy before escaping. By comparing the orbits in strongly deformed fields due to the KHI development, with the textbook mirror-drift orbits resulting from our initial configuration, we identify effects due to current sheets formed in the DMLR process. We do this
in various representative stages during the DMLR development.
\end{abstract}


%
%

%


%
%
%
%


\section{\label{sec:intro}Introduction}
A key mechanism in solar wind (SW)/magnetosphere interaction, affecting all magnetized planets in our heliosphere, is the Kelvin-Helmholtz instability (KHI), usually developing on the flanks of the 
magnetosphere~\cite{dungey1968waves,nature2004hasegawa}. The KHI is a fundamental fluid instability driven by shear flow. In a magnetosphere, the magnetopause represents the interface where the shocked SW plasma flows past more static magnetospheric plasma,  and especially when the magnetic field is oriented almost perpendicular to the shear flow direction, the KHI can develop quite naturally. Using the Magnetospheric Multiscale Mission (MMS) and THEMIS spacecrafts, the KHI was recently observed simultaneously at both the dawn and dusk flanks of the Earth~\cite{Luetal2019}, behaving in an almost quasi-symmetric fashion. The KHI and reconnection processes associated with its nonlinear development can determine the efficiency of the plasma transport between SW and magnetospheric regions, and is therefore a topic of intense ongoing research.  A detailed analysis of 7 years of THEMIS data revealed that the KHI may well have been operative during 19\% of this time interval~\cite{Kavosi2015}.

KHI in magnetized plasmas has extensively been studied both theoretically and numerically, especially in a single fluid magnetohydrodynamic (MHD) viewpoint. Using linear MHD theory, one can quantify linear growthrates and eigenfunctions for KHI modes in stationary equilibrium configurations~\cite{Miura1982}. For highly idealized configurations, analytic expressions for the KHI linear growth rate are available in textbooks, e.g. in \citeA{Chandra1961}. When the flowing MHD equilibrium has non-trivial, sheared magnetic field variations, in combination with sheared flow and rotational profiles, precise quantifications of the eigenfrequency-eigenfunction variations on basic equilibrium parameters like the plasma beta or the flow Mach number are actually far from trivial. There, the recently developed spectral web~\cite{Goedbloed2018,Hansbook3} approach represents a new means to compute all complex eigenfrequencies accessible to a magnetized force-balanced state in motion. 

To address how KHI leads to plasma transport or reconnection, fully nonlinear numerical simulations are used.  Idealized 2D setups in MHD identified important differences between KHI development in parallel versus antiparallel magnetic field orientations~\cite{Keppens1999}, or studied the effects of vortex coalescence in extended shear layers~\cite{Baty2003,Nakamura2008}. A comparative study between MHD, Hall-MHD, 2-fluid, hybrid and full kinetic treatments of the KHI was performed by~\citeA{Henri2013}, where the large-scale behavior was found to be properly described by fluid treatments. Fully three-dimensional kinetic treatments for the KHI have been performed as well~\cite{Nakamura2013}, stressing how KHI vortices drive strong current sheets where tearing causes magnetic flux ropes to form, which get advected and finally merge in the vortex flows. 

Simulations of local Kelvin-Helmholtz evolutions applicable to magnetopause configurations have been performed by many authors, with parameters that mimic conditions at Saturn~\cite{Delamere2011,Maetal2015}, Earth~\cite{Nykyri2001}, or the ionopause of Venus~\cite{Li2019}. Three-dimensional studies of local KHI evolutions at the Earth's magnetosphere have gradually realized the importance of the higher-latitude double reconnection process, that is triggered by the KHI equatorial development. During northward interplanetary magnetic field (IMF) conditions, local 3D MHD studies showed that an efficient plasma transport can be established~\cite{Maetal2017}. In this double mid-latitude reconnection (DMLR) process \cite{epl2012faga,epl2014faga,borgogno2015pop,faganello2017jpp}, the KHI self-consistently creates configurations liable to magnetic reconnection and subsequent particle acceleration. The goal of the present manuscript is to start from our recently performed, 3D resistive and Hall-MHD simulations of the DMLR process~\cite{pop2017leroy}, and determine various aspects related to how charged particles behave in these configurations. Our study covered multiple wavelengths of the most unstable KHI mode, and was able to follow the KHI development into its coalescing regime. 

In global magnetospheric models, it remains numerically challenging to capture local shear-flow related details, although KHI development has been demonstrated for a northward IMF condition at Earth in~\citeA{Guoetal2010},
in global kinetic hybrid simulations for Mercury~\cite{Paral2013} where the vortices are mainly advected to the dusk side, and recently also in the case of the Jovian magnetosphere, where dawn-dusk assymmetries play a crucial role~\cite{Zhangetal2018}. As we are interested in how the details of the KHI development determine local particle orbits, we will use the fully 3D, local box DMLR configurations representative of the magnetopause variation during KHI, and study particle motion in specific phases of the MHD evolution. 
Since the MHD approach no longer contains information on the intricate particle trajectories, a first step to identify e.g. trapping sites for particle populations is to use test particles in given MHD fields. 

The test particle approach is known from textbook treatments of charged particle motion in given electric and magnetic field configurations, introducing drifts associated with $\mathbf{E}\times\mathbf{B}$, $\nabla \mathbf{B}$ and magnetic field curvature~\cite{Sturrock1994,GurnettBat,Hansbook3}. In the context of magnetospheric physics, one can adopt semi-empirical prescriptions for planetary magnetic fields, and solve the (relativistic) equation of motion in them to discuss aspects related to trapping, orbit geometries, loss cone aspects, etc. This was e.g. done by~\citeA{Walsh2013}, where a model magnetic topology for the magnetosphere of Mercury was adopted, and the dipole offset ensured that north-south asymmetric loss cones form, predicting more particle precipitation in the southern hemisphere. An analysis of particle motion in the Jovian magnetosphere~\cite{Mahjouri1997}, modeled by a superposition of a dipole field and a parametrized current sheet disc fitted to Pioneer 10 data, could identify escape probabilities through the current disc. Using magnetic field models based on Voyager measurements, \citeA{Birmingham1982} performed guiding centre approximation (GCA) simulations of particle mirroring within Jovian and Saturnian fields, and found the stronger effects in the Jovian case due to this magnetodisc current. These examples all exploit model, global time-independent field configurations, while in this work we target particle motion aspects related to KHI development, for which the local MHD simulations provide detailed field topologies.

The spatial and temporal scales involved for the DMLR at the Earth's magnetosphere led us to perform test particle simulations using the GCA 
framework~\cite{northrop1963adabatic,astro2018ripperda}. The particles are embedded in a background thermal plasma, using MHD field 
configurations from previous simulations that represent the ion dynamics in a single fluid viewpoint~\cite{pop2017leroy}. Test particle 
trajectories are determined by the electromagnetic fields, without feedback of the particles on those fields. In a plasma where the gradient of the magnetic field is 
larger than the gyration radius of the particle, the gyromotion can be averaged over and the GCA can be applied, allowing 
for simplified equations for the particles. Details of the method and justifications of its use will be presented in Section \ref{subsec:part}. 
Using the instantaneous electromagnetic fields from the MHD solution, the electric field has 
spatially varying, parallel components due to the resistive processes incorporated in the MHD runs. By ignoring the time-variability of the magnetic 
fields, the magnetic field will only transfer parallel to perpendicular kinetic energy since the Lorentz force can not do work. Still, as we model particle 
trajectories in spatially complex configurations that result from the DMLR evolution, the combined effects of gradient and curvature drifts are properly 
represented in the GCA approach, and are vital to understand the complex particle trajectories. Various tests demonstrating the excellent agreement between e.g. GCA and fully gyrating particle motion in mirror configurations and dipole fields have been demonstrated in~\citeA{astro2018ripperda}.

The manuscript is organized as follows. In Section \ref{sec:numset} we present the MHD model and its numerical treatment along with the initial conditions 
devised to simulate the particle trajectories inside the DMLR. 
Section \ref{sec:accel} presents the different trapping and particle dynamics scenarios which can take place in the DMLR configurations. Section~\ref{sec:conclu} sums up our findings and provides an outlook for further work.

\section{\label{sec:numset}Physical and numerical setup}

The test particle simulations presented in this article were realized using MPI-AMRVAC, the parallelized Adaptive Mesh Refinement Versatile Advection Code~\cite{keppens2012parallel,ajss2014porth,Xiaetal2018}. The resistive and Hall-MHD module 
(extension of MHD applicable to phenomena occurring on length scales shorter than the ion inertial length, and time scales shorter than the ion cyclotron period) 
has been tested and validated by \citeA{ajss2014porth} and the test particle module has been introduced and further applied by~\citeA{ripperda2017mnras,ripperda2017mnras2,astro2018ripperda}. This section will recall relevant equations and initial settings directly related to our study.

\subsection{\label{subsec:mhd}The MHD setup and evolution}
The test particles are to be evolved in background electromagnetic fields extracted from full 3D Kelvin-Helmholtz evolutions, presented in~\citeA{pop2017leroy}. In order to characterize 
how the large-scale Kelvin-Helmholtz instability can impact the trajectories of the particles, three different simulation times are selected 
representing three stages of the instability. A reference case will be established, by analysing possible particle trajectories in the initial $t_{MHD}$=0 snapshot, where $t_{MHD}$ is the normalized Alfv\'en time as used in the MHD simulation. These are the initial conditions of our simulation, where the flanks of the magnetopause are unperturbed. Then, the influence of the 
instability will be investigated by injecting particles with the same initial conditions into snapshots at $t_{MHD}$=400 (this corresponds to the rolling-up of the waves) and $t_{MHD}$=600 
(when we have a fully non-linearly developed Kelvin-Helmholtz instability). 

Though a complete description can be found in the DMLR parameter study~\cite{pop2017leroy}, 
the main features of the MHD evolution will be recalled here for clarity. The box size is chosen as $L_x$=70, $L_y$=188, $L_z$=377 
(with $x\in[-40,30], y\in[-L_y/2,L_y/2], z\in[-L_z/2,L_z/2]$), all given in ion inertial lengths $\delta_i=c/\omega_{p}$ ($\approx$100 km in near Earth conditions). The grid resolution in the simulations is $200^3$. The $z$-direction corresponds to latitude where $z=0$ is equatorial, the $y$-direction follows the magnetopause interface 
where the shear flow instability will develop, while the $x$-direction points across the shear layer (i.e. along $x$ it goes from solar wind plasma into magnetospheric plasma conditions).

From $\delta_i=100$ km, we find that the ion (and electron, due to charge neutrality) number density $n_i=5.16 \,\mathrm{cm}^{-3}$, which is a typical value near the magnetopause~\cite{Hubert1998}. The MHD cell size is then $\Delta z=180 \,\mathrm{km}$, or of the order of the ion inertial length. The full vertical box extent of $377\delta_i$ should be compared to a typical field line length for a dipolar Earth field line reaching out to the magnetopause. Taking an $L$-shell of 10 to 15 (with $L$ the equatorial standoff distance in Earth radii), an estimate consistent with equating solar wind ram pressure with dipolar magnetic pressure, our vertical box size is about one fifth to one tenth in length of a pole to pole field line. Here, we used that a dipolar field line measures about $3L$ in length.

The simulation is initiated with analytical fields derived from the 
solution of a simplified Grad-Shafranov equation, that is invariant in the $y$-direction, and describes an initial equilibrium, such that 
the $\nabla \cdot \mathbf{B}$ = $0$ constraint is satisfied \cite{faganello2012ppcf}. The magnetic fields are derived from a vector potential 
\begin{equation}
A_y (x,z)=(1/2)\left(4x/3+L_x/(2\pi)\sinh\left(2\pi x/L_z\right)\cos\left(2\pi z/L_z\right)\right) \,, \end{equation} in the usual fashion 
$B_x$=$-\partial A_y/\partial z$, $B_y$=0, $B_z$=$\partial A_y/\partial x$. The initial flow configuration has $V_x$=0, $V_y$=$(M_A/2)\tanh(A_y/L_u)$, $V_z$=0, i.e. the sheared flow is along the $y$-direction, with its main (shear) variation along $x$. 
The velocity field is destabilized by incompressible perturbations 
\begin{equation}
\delta A_y=\epsilon\sum{_{m=1}^6}\Big(\cos(2\pi mx/L_y+\phi(m))/m\times\exp{-(x/2L_u)^2}\times\exp{-(z/2L_u)^2}\Big) \,,\end{equation}
with amplitude $\epsilon$=0.05 and $\phi(m)$ a set of random phases. The density follows the same profile as the velocity, i.e.
\begin{equation}
\rho(x,z)=(1/2)(\rho_c+1)+(1/2)(\rho_c-1)\tanh(A_y/L_u)\,, \end{equation} 
where we introduce a realistic density contrast between solar wind (magnetosheath) and magnetospheric plasma.
For the reference run, $\rho_c$=4.7, leading to an initial density contrast of 3.7 with the shocked solar 
wind magnetosheath plasma presenting a higher density than the magnetosphere. This density increase is consistent with measurements~\cite{Hubert1998}. The pressure is set as half the plasma beta parameter $\beta$=0.71 and is 
constant throughout the domain to respect the initial equilibrium. The other significant parameters set at the initialization of the simulations are the Alfv\'en 
Mach number $M_A$=1, the sonic Mach number $M_c$=1, the half-width of the shear layer $L_u$=3 and the polytropic index $\gamma$=5/3.
The resistivity employed in the resistive MHD run is chosen as $\eta$=0.001 to represent a nearly collisionless, weakly diffusive plasma. The initial analytical magnetic field is potential (i.e. $\mathbf{J}=\mathbf{0}$), ensuring that the initial electric field is exactly orthogonal to the magnetic field, since 
\begin{equation}
\mathbf{E}(t_{MHD}=0)=\left[-\mathbf{v}\times\mathbf{B}+\eta\mathbf{J}\right]\mid_{t_{MHD}=0}=\left[-\mathbf{v}\times\mathbf{B}\right]\mid_{t_{MHD}=0} \,,
\end{equation}
This is consistent with the collisionless nature of space plasmas, where parallel electric fields are rather rare. In the later MHD snapshots, a small parallel electric field does develop, but we will see that it only plays a minor role. The maximal electric field component is always in the $x$-direction (across the shear flow and the magnetopause, as $\mathbf{v}$ is mostly along $y$, while $\mathbf{B}$ is dominated by its $z$-component), and reaches order $100-1000$ nV/m. The $E_y$ and $E_z$ components are usually one to two orders of magnitude smaller. The parallel electric field is thus in practice restricted to the KHI induced current sheets, where some anomalous resistivity (corresponding to our $\eta=0.001$) may be a reasonable first proxy.

\begin{figure}
\centerline{\includegraphics[width=\textwidth]{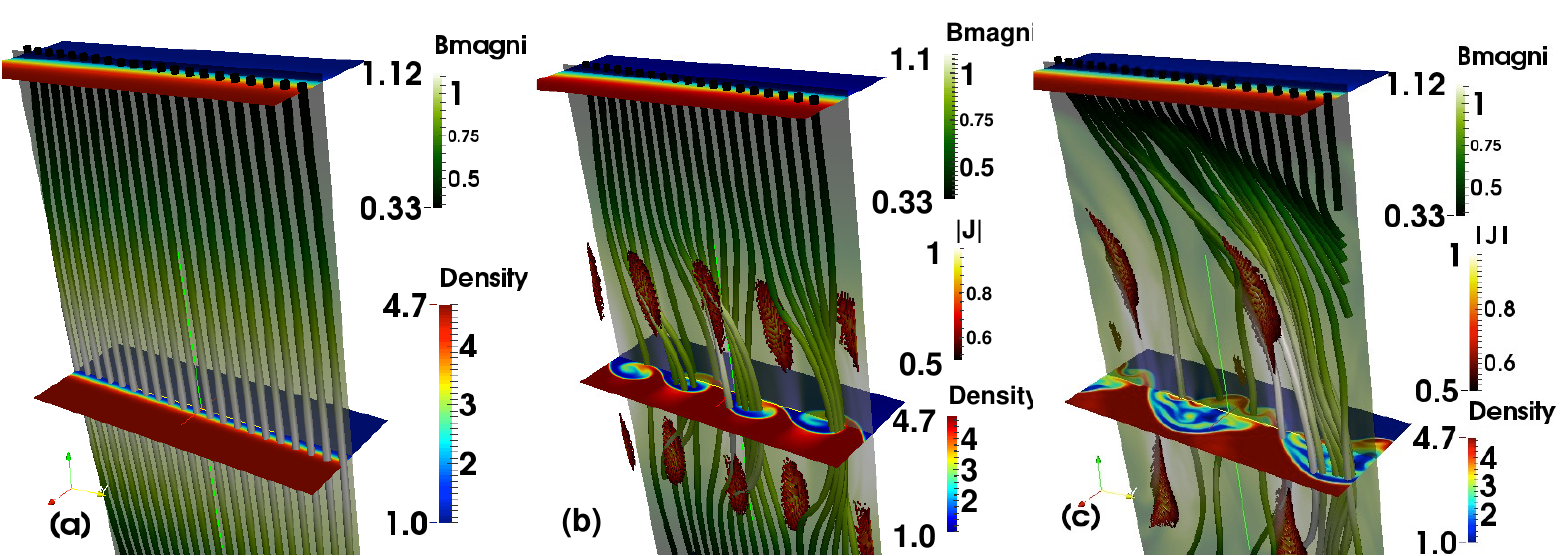}}
\caption{\label{fig:DMLR_background}Time evolution of the MHD simulation. Slices in the horizontal $x-y$-plane present the density at different latitudes. 
Magnetic field lines are coloured with magnetic field magnitude (in green), also represented by a vertical slice in the $y-z$-plane. Points where the current magnitude 
is above 0.5 NU are displayed in shades of orange. Fig.(a) $t_{MHD}$=0, Initial configuration. Fig.(b) $t_{MHD}$=400, Rolling-up. Fig.(c) $t_{MHD}$=600, Mixing layer 
and strong current.}
\end{figure}

These initial conditions induce a differential advection of the field lines as the flow evolves. Indeed the `high-latitude' areas (top and bottom edges of the periodic box in the vertical $z$-direction) 
are more stable with respect to the KHI than the domain closer to the $z$=0 plane, due to variations of the velocity and magnetic field magnitude. This is a local representation for the actual conditions at the magnetopause flanks, where the more equatorial regions are most liable to Kelvin-Helmholtz roll-up.
Thus vortices emerging near the $z$=0 plane locally advect the field lines while the part that is further away remains unperturbed for a longer time.
The snapshots in Fig.~\ref{fig:DMLR_background} show the selected background configurations from the MHD simulation in normalized units (NU). 
The horizontal slices are coloured with the density (blue to red colourscale) and the current magnitude is displayed in surfaces in shades of orange. 
The magnetic field is represented by a vertical slice in $y-z$-plane and by field lines in shades of green. The simulation box presents a symmetry with regards to the equatorial 
plane. At $t_{MHD}$=0 (Fig.~\ref{fig:DMLR_background}a) the initial configuration displays slightly hour-glass shaped magnetic field lines, where the field is nearly straight as it is dominated by the 
$z$-component with its maximum value at $z$=0 and decreasing symmetrically towards the high latitudes. 
In the rolling-up stage ($t_{MHD}$=400, Fig.~\ref{fig:DMLR_background}b), the 
magnetic field lines relax around the crests of the waves, called spines of the KHI by~\citeA{2014jgramaotto1}, while they are compressed in the lower 
density (light blue) area of the waves. As a consequence of the creation of vortices, some field lines are starting to intersect and reconnect
at the latitudes where the flow turns from KHI unstable to KHI stable (latitude where the field lines turn from straight to bent). In this magnetic 
reconfiguration, current sheets start to form at mid-latitudes, as seen in the middle panel both above and below the equatorial midplane. In the last snapshot 
($t_{MHD}$=600, Fig.~\ref{fig:DMLR_background}c), the rolled-up vortices have merged, inducing fewer but more extended current sheets that also attain a higher current magnitude. 
The intersecting field lines obtain an even stronger deformation and curvature due to the differential advection. This process keeps increasing
and reconnection sites are generated in a doubled fashion: below and above the equator. Field lines that were previously connected to the magnetosphere can thereby become linked to the solar wind, and vice versa, due to 
reconnection. This is called the DMLR phenomenon \cite{epl2012faga,epl2014faga}. It occurs around the current sheets for as long as 
the KHI vortices keep on braiding the field lines and the resulting effect is an almost regular and instantaneous exchange of matter and energy between the two
regions. An extensive discussion of the fluid processes occurring, along with a detailed parameter study, can be found in our previous work~\cite{pop2017leroy}. 

\subsection{\label{subsec:part}Particle treatment and initial conditions}

In order to integrate the trajectories of the test particles, the GCA~\cite{northrop1963adabatic} will be used in this paper. Taking into account the length scales and field magnitudes present in the MHD simulations, this 
approximation accurately describes the particle trajectories as characterized by their charge $q$ and (rest) mass $m_0$. Indeed, as long as their gyroradius $r_g$=$(m_0v_{\perp})/(|q|B)$ and 
gyroperiod $P_g$=$2\pi/\Omega$=$2\pi m_0/(|q|B)$ are much smaller than the lengthscale of the field gradients and than the characteristic oscillation periods of the background 
electromagnetic fields, respectively, their motion can be decomposed into the trajectory of their guiding centre and a fast gyration around this centre, symbolized by a constant 
magnetic moment. The GCA approach leads to five equations describing the evolution of the guiding centre position vector $\mathbf{R}$, the parallel momentum $\textit{p}$=$m_0 \gamma v_{\parallel}$ 
and the magnetic moment $\mu$=$m_0 \gamma^2 v_{\perp}^2/(2B)$, with Lorentz factor $\gamma$=$1/\sqrt{1-v^2/c^2}$ and $c$ the speed of light. The relativistic 
form \cite{vandervoort1960relativistic} of the GCA equations will be used here for completeness, where they read as
\begin{eqnarray}
\frac{\partial \big(m_0 \gamma^2 v_{\perp}^2/(2B)\big)}{\partial t} & = & 0, \label{perp}  \\
\frac{\partial(m_0 \gamma v_{\parallel})}{\partial t} & = &m_0\gamma\mathbf{v}_E\cdot\Big(v_{\parallel}(\mathbf{b}\cdot\nabla)\mathbf{b}+(\mathbf{v}_E\cdot\nabla)\mathbf{b}\Big)
+ qE_{\parallel}-\frac{\mu}{\gamma}\mathbf{b}\cdot\nabla(\kappa B), \label{parallel} \\
\frac{\partial\mathbf{R}}{\partial t} & = & \frac{(\gamma v_{\parallel})}{\gamma}\mathbf{b}+\frac{\mathbf{b}}{\kappa^2}\times\Bigg\{-\kappa^2c\mathbf{E} \nonumber \\ & &
+\frac{cm_0\gamma}{q}\Big(v_{\parallel}^2(\mathbf{b}\cdot\nabla)\mathbf{b}+v_{\parallel}(\mathbf{v}_E\cdot\nabla)\mathbf{b}+v_{\parallel}(\mathbf{b}\cdot\nabla)\mathbf{v}_E 
+(\mathbf{v}_E\cdot\nabla)\mathbf{v}_E\Big)
\nonumber \\ & &
+\frac{\mu c}{\gamma q}\nabla(\kappa \mathbf{B})+\frac{v_{\parallel}E_{\parallel}}{c}\mathbf{v}_E\Bigg\}, \label{gyro}
\end{eqnarray}
where $v_{\parallel}$ and $v_{\perp}$ are the velocities parallel and perpendicular to the magnetic field respectively, and 
$\mathbf{b}=\mathbf{B}/B$ is the unit vector in the direction of the magnetic field.
The drift velocity $\mathbf{v}_E$=$\mathbf{E}\times\mathbf{B}/B^2$ is the drift induced by the $\mathbf{E}\times\mathbf{B}$ field and $\kappa$ is its Lorentz factor. The electric field $\mathbf{E}$ has a component parallel to the magnetic field denoted by $E_{\parallel}$, as the simulation used a resistive MHD prescription. These governing equations are solved using a fourth-order Runge-Kutta scheme with adaptive time-stepping and the field values are interpolated from the original $200^3$ grid to the particle positions to first order, after 
being scaled to CGS units from the normalized MHD simulations.

Since we are interested in the effects of the perturbations happening at mid-latitude in the MHD snapshots, particles are initiated within a narrow 
band of the following dimensions $x\in[2,8], y\in[-L_y/2,L_y/2], z\in[25,85]$, defined to encompass the more intense part of the current sheets present at 
$t_{MHD}$=400 and $t_{MHD}$=600. Their velocities are initiated following a Maxwellian distribution and with random directions. 
With typical values of the SW/magnetosphere interface~\cite{oieroset2008themis} $v_{\perp}\sim40$km/s and $B$=$40$nT, the gyroradius for electrons 
is of the order $r_g \sim 10$ m, while the MHD grid cell size is 180 km, of the order of $\delta_i=100$ km. So even if the particles are accelerated to 
non-thermal velocities, the gyroradius may increase up to several orders until the GCA fails. We initialize electrons with a Maxwellian around a thermal velocity $U_0\sqrt{2 m_p/m_e}$, where $U_0$ is an Alfv\'en speed of order 650 km/s. The factor $\propto \sqrt{m_p/m_e}$ accounts for the higher mobility of electrons, and the value of the Alfv\'en speed of 650 km/s, combined with the ion number density $n_i=5.16 \,\mathrm{cm}^{-3}$ from $\delta_i=100$ km, is consistent with a magnetic field of order $B$=$65$ nT. These ${\cal{O}}(10 \,\mathrm{nT})$ fields are roughly consistent with $L$-shell 10 field strengths, as the equatorial magnetic field $B\approx B_E/L^3$ with $B_E=0.3 \,\mathrm{G}$. The typical particle velocities become order $v_p\approx 2 \times 10^4 \,$ km/s, or in the keV range. For typical keV electrons, the actual bounce periods due to mirroring in the Earth dipolar field at both poles become order 10-20 seconds. A particle traverses our $L_z$ box in order 1 second, which is a fraction of the bounce period, consistent with the difference between the vertical box extent and the true pole-to-pole dipolar field line length.

The typical timestep for the evolution of the variables related to the test particles is much smaller than the MHD 
timestep, so the MHD fields can safely be considered static, and we will denote the employed particle time with $t_p$. We note that when MHD fields are considered static, effects like particle acceleration in converging magnetic mirrors (i.e. Fermi processes) are not possible, but all gradient and curvature drift effects are properly captured. We evolve electrons, because their gyroradius is smaller and thus the GCA holds better for electrons than for heavier ions. 
Moreover, since electrons are lighter and faster, they will traverse a longer trajectory in the same physical time, allowing them to cover more of the physical domain and 
thus giving us access to the various drift processes that can be encountered. The ions will be considered as a single fluid as represented by the MHD 
background fields, thus governing the evolution of the test particles.  
As long as we focus on orbit aspects from particles that move fast relative to the field temporal variations, 
the test approximation is fully justified. 
As we find that our representative particles actually do not reach non-thermal velocities, the non-relativistic form of the above GCA equations will be used in our further analysis. We will typically advance the particles to $t_p$=20, sufficient to demonstrate intricate particle orbits traversing our entire box. A note of caution is in order: a true statistical treatment for many particles is possible in principle, but in practice suffers a severe drawback resulting from the local box approach of our 3D simulation: many particles will escape through the back and front of the domain as our boundary 
conditions in the $x$-direction are open. Our initial conditions result in an hourglass shape for the magnetic field lines, with the narrowest and strongest) part located 
on the equatorial plane. Therefore, several field lines connect a $z>0$ part of the front (or back) edge to a $z<0$ part. Particles that are initially on these fields lines rapidly exit the simulation box. An experimental run with 10000 particles initially thus rapidly reduces to retain only about 2000 particles at $t_p=2$. 

The initial hourglass magnetic field setup has its strongest field at the equator, opposite to the large scale pole-to-pole variation of the dipolar field in which our local box would be embedded. However, the initial variation from the middle (equatorial) to top and bottom regions in our box amounts to a change by only a factor of two in field strength, and the KHI development introduces significant changes in all components, with the introduction of localized strong current regions. By comparing the orbits in the later KHI phases with the fiducial orbits found for the $t_{MHD}=0$ snapshot, we identify aspects that would be generic to KHI variations alone.

As a method to understand how the particle population is driven by the MHD configurations, we will typically quantify the parallel acceleration terms conform 
the parallel magnetic gradient or mirroring term 
$\mathbf{b}\cdot\nabla(\kappa B)$, the parallel resistive field acceleration term $qE_{\parallel}$,  
and the curvature terms $\mathbf{v}_E\cdot[(\mathbf{v}_E\cdot\nabla)\mathbf{b}]$,  and 
$\mathbf{v}_E\cdot[(\mathbf{b}\cdot\nabla)\mathbf{b}]$. Note that these are all scalar quantities, which contribute to parallel 
acceleration as expressed in the GCA equations. 
In the $t_{MHD}$=0 snapshot, we will find that the electric field acceleration and parallel gradient terms are slightly dominant. This is consistent with Fig.\ref{fig:DMLR_background} where the MHD background acts as magnetic trap 
with the field line's density and magnitude increasing toward the equatorial plane. The particles can be expected to bounce around a magnetic trap and to be
accelerated in the middle by an electric field, while slowly drifting. The curvature terms are already influencing the motions  because of the hourglass shape of the magnetic field 
discussed earlier. In the later KHI stages, we expect that the curvature 
terms are largely dominant, coherent with Fig.~\ref{fig:DMLR_background} displaying twisted field lines. In these configurations, the particles can be 
bouncing in the mirror trap while following strongly curved trajectories with drifts caused by the large magnetic gradients.
Since the particles in the DMLR setup here are mostly following periodic orbits, the forces they actually experience display periodicity and local variations in 
magnitude as well. This calls for a particle-by-particle analysis in order to clearly assess which processes determine their individual orbit scenarios.

\section{\label{sec:accel}Individual particles trajectories and acceleration mechanisms}

Particles evolving in the electromagnetic background fields from KHI evolutions are seldom subject to one or another acceleration mechanism in 
a simple manner. Analysis of individual particle trajectories can inform us on the precise ways and times the different momentum contributions act
and which mechanisms could accelerate particles. 
It is reminded here that particles that cross the top or bottom edges of the domain re-enter on the opposite edge because we handle this boundary periodic. In reality, particles leaving the equatorial regions northward can evidently not come back in the south. Our periodic boundary 
conditions in both $y$- and $z$-direction force them to remain in the simulation box, consistently with the periodic boundaries applied in the MHD study. Since our 
focus of this study is on processes taking place around the equatorial and mid-latitude region, we can imagine their forced repeated visits to the equatorial site as sampling 
particles launched into the simulation with the same direction but different initial positions. As our vertical box extent is a fraction of about one tenth of the real dipolar field line extent, we must resort to reinjecting the particles in the $z$-direction. By treating the particle orbits as $z$-periodic, we interpret this as an effective means to sample multiple particle trajectories that would in reality be on bouncing orbits.

\begin{figure}
\centerline{\includegraphics[width=\textwidth]{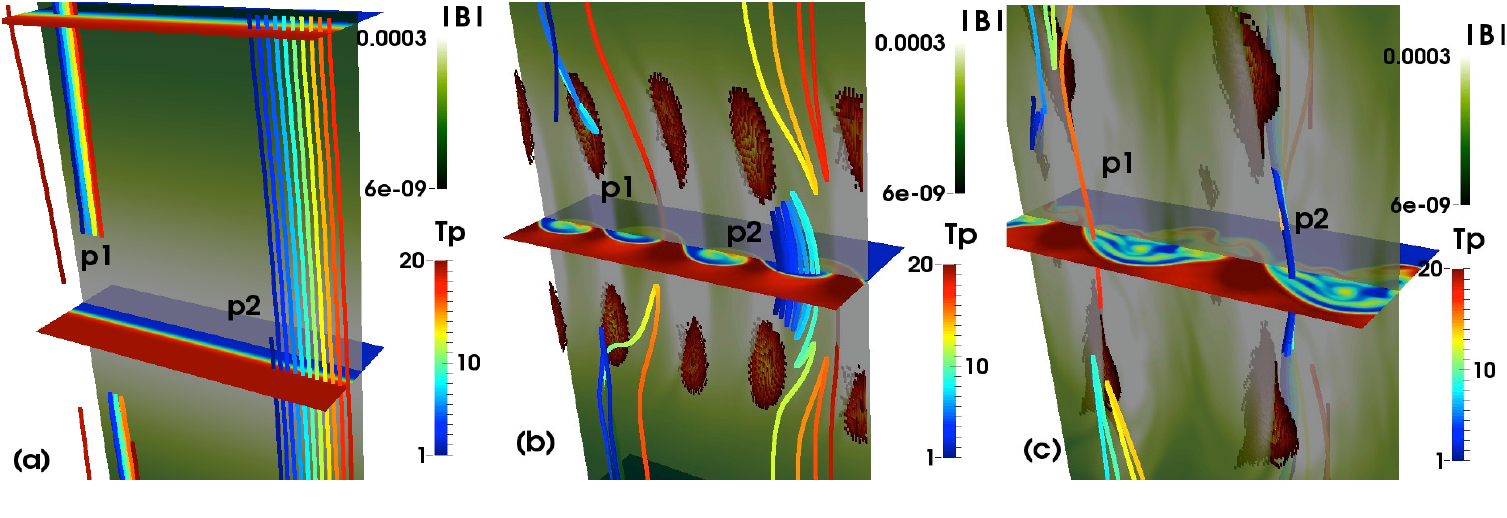}}
\caption{\label{fig:traj10p}Trajectories for particles p1 and p2, coloured with time for the 3 configurations. All other colours have the same meaning as 
Fig.~\ref{fig:DMLR_background}.}
\end{figure}

Figure~\ref{fig:traj10p} displays trajectories of two particles, coloured with time. The chosen particle trajectories are typical for our study, and between the three MHD snapshots, each particle displayed has the same initial position and velocity across the simulations.
As has been hinted before, the trajectories are different between the $t_{MHD}$=0 simulations and the others. 
Fig.\ref{fig:traj10p}.(a) shows the only two possible types of trajectories for the $t_{MHD}$=0 state. The first kind of particles like p1 are trapped in mirroring trajectories away from  
the equatorial plane, where complete mirroring occurs at a finite distance from the equatorial plane. Those particles bounce back and forth in the 
magnetic bottle created by the larger magnitude of the magnetic field around $z$=0 (note that the larger magnitude near the equator is enforced in our initial condition, and may seem opposite to a pure dipole variation where the field is weakest there), and their individual excursion in the $z$-direction is determined by their initial 
energy. The second kind like p2 are streaming across the domain when their energy is sufficient to escape the magnetic trap. 
Note also the small drift in the direction perpendicular to the (mainly vertical) magnetic field as time progresses, due to the $\mathbf{E}\times\mathbf{B}$ contribution. 
This is coherent with the initial topology constituted by a 
large $z$-component of the magnetic field with areas of larger values of the $x$-component around the mid-latitudes, creating the mirror trap. These results for the $t_{MHD}$=0 case just serve as a reference to compare with in the later more complex stages.

Figure~\ref{fig:traj10p}.(b) illustrates how mirroring particle trajectories are modified by bending of the magnetic field lines. The curvature terms are now 
more dominant and those particles can drift much further to other field lines. It also introduces particles, like p2,  temporarily trapped between 
the current sheet latitudes. As can be seen from the trajectories of both particles, it is easier to penetrate the equatorial region when 
passing between the current sheet regions, which corresponds to regions of lower magnitude of the magnetic field. Particles coming from the exterior can get trapped 
in the equatorial region, to finally escape (like p2) if they drift onto an easier path through the magnetic trap.
Figure~\ref{fig:traj10p}.(c) completes the study for later times, by showing that particles can get trapped in smaller regions close to the current sheet, like found for p1 at 
the beginning of the simulation (i.e. during the blueish section of its path at the top left of the panel). It demonstrates how locations where a particle bounces back or passes through the magnetic trap can be very close. It also shows 
that overall statistics will hide specifics of trajectories, showing here how this time p1 is trapped and escapes while p2 passes through or mirrors outside the mid-latitude. 

In summary, when the magnetosphere is in a quiet configuration such as supposed by the $t_{MHD}=0$ snapshot, particles traverse a region where the magnetic field 
lines are compressed near the equatorial plane ($z$=0). If the Kelvin-Helmholtz instability develops and starts twisting the field lines at the interface, the conditions 
for more complex trajectories emerge and the various electromagnetic gradients influence the kinetic energy evolution (and its repartition over parallel and 
perpendicular energy) of individual particles. The particles encounter 
larger magnetic gradients, electric fields in larger and more frequent spots, leading to the increase of the drift magnitude. Particles can also now temporarily or 
more permanently be caught in a localized magnetic trap near the equatorial region, set up by the KHI process.

\subsection{\label{subsec:dcacc}Accelerated particles}

In this section, the different types of trajectories will be examined more quantitatively, to identify the mechanisms affecting the particle's gain or loss of kinetic energy. 
The time evolution of their position, kinetic energy and the electromagnetic fields they encounter will 
be compared against the different terms directly affecting the kinetic energy variations. Indeed, combining Eqs.~(\ref{perp}) and (\ref{parallel}), an expression for 
the rate of kinetic energy change can be derived (from~\citeA{astro2016zhou} with additions) :
\begin{eqnarray}
\frac{\partial E_k}{\partial t} & =&  \frac{m_0}{2}\frac{\partial(\gamma v_{\parallel})^2}{\partial t}+\frac{m_0}{2}\frac{\partial(\gamma v_{\perp})^2}{\partial t}, \label{Ek}\\
\frac{m_0}{2}\frac{\partial(\gamma v_{\perp})^2}{\partial t} & = & \mu v_{\parallel}(\mathbf{b}\cdot\nabla) B+\mu (\mathbf{v}_E \cdot\nabla)B, \label{Eperp}\\
\frac{m_0}{2}\frac{\partial(\gamma v_{\parallel})^2}{\partial t} & = & q\gamma v_{\parallel}E_{\parallel}+m_0(\gamma v_{\parallel})^2 \mathbf{v}_E\cdot[(\mathbf{b}\cdot\nabla) B]\nonumber\\ & &
+m_0 \gamma^2 v_{\parallel} \mathbf{v}_E\cdot[(\mathbf{v}_E \cdot\nabla)B]-\mu v_{\parallel} \mathbf{b}\cdot\nabla(\kappa B) \,. \label{Epar}
\end{eqnarray}
It is verified for all our particles studied here that $\gamma \sim 1$ and hence $\kappa \sim 1$. Equation~(\ref{Eperp}) and Eq.~(\ref{Epar}) can thus be substituted into Eq.~(\ref{Ek}) to obtain 
a final expression for the kinetic energy evolution, which essentially retains four terms:
\begin{eqnarray}
\frac{\partial E_k}{\partial t} & = & q v_{\parallel}E_{\parallel}+\mu (\mathbf{v}_E \cdot\nabla)B+m_0 v_{\parallel}^2 \mathbf{v}_E\cdot[(\mathbf{b}\cdot\nabla) B]\nonumber\\
& &
+m_0 v_{\parallel} \mathbf{v}_E\cdot[(\mathbf{v}_E \cdot\nabla)B] \,. \label{Ekinevol}
\end{eqnarray}
This equation confirms that even if the parallel magnetic gradient $\mathbf{b}\cdot\nabla B$ can act on $v_{\parallel}^2$ and $v_{\perp}^2$ variations, it 
has no effect on the total kinetic energy, in accord with the fact that magnetic fields do no work. The remaining terms express the resistive electric field $q v_{\parallel}E_{\parallel}$ or $\textit{resist}$, 
the perpendicular magnetic gradient $\mu (\mathbf{v}_E \cdot\nabla)B$ or $\textit{gradb}$, and the curvature terms contributions respectively (identified as $\textit{b.curvb}$ and $\textit{ue.curvb}$ in the following figures). 
In what follows, the contributions of these scalar terms to the kinetic energy evolution are shown, and thereby we quantify acceleration effects. 
Note that not all terms lead to field-aligned acceleration, e.g. the $\textit{gradb}$ term affects the perpendicular velocity component.

\begin{figure}
\centerline{\includegraphics[width=0.8\textwidth]{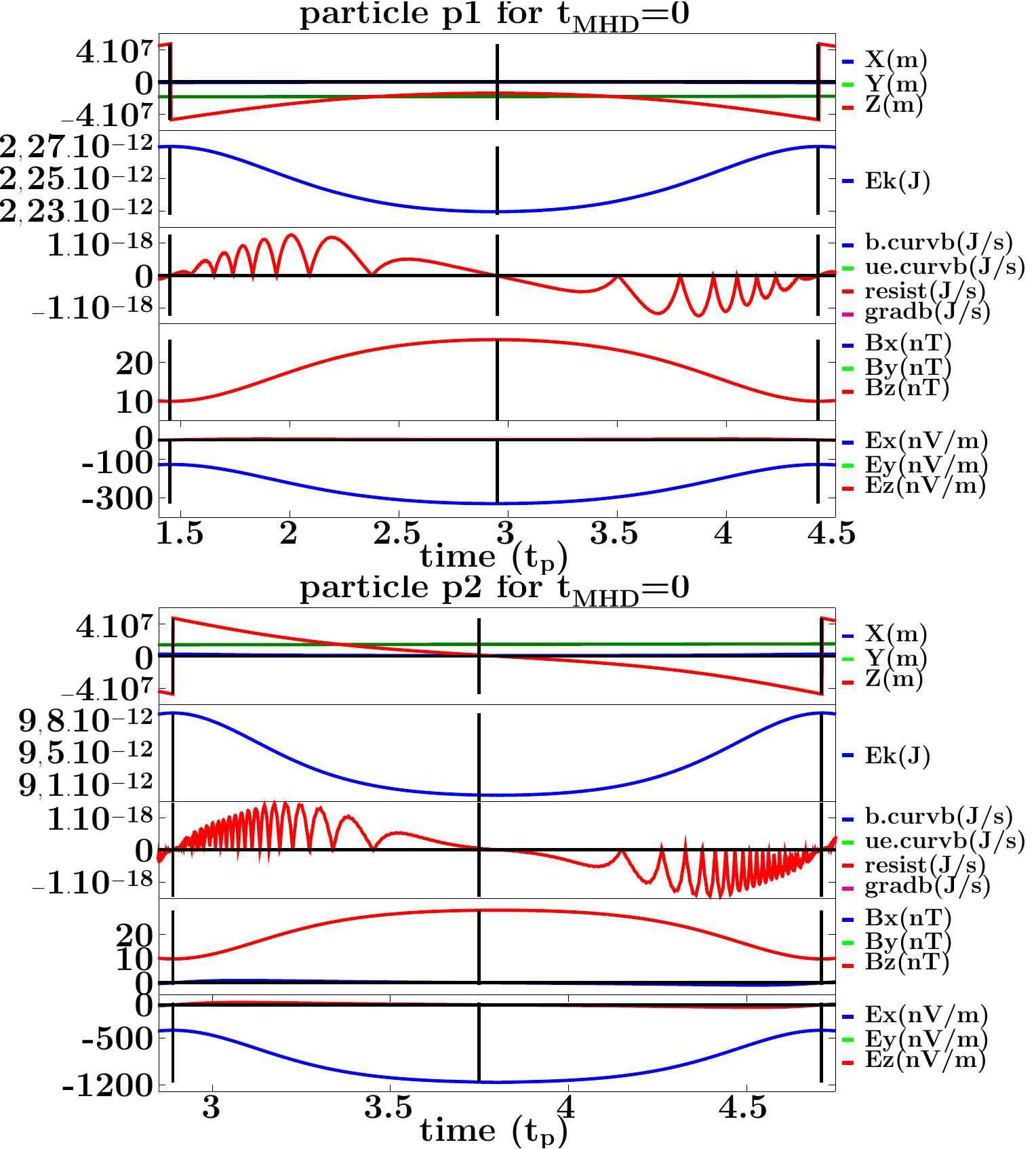}}
\caption{\label{fig:diag_part_t00}Positions, kinetic energy, $E_k$ contributions and electromagnetic fields encountered for 
particles p1 (top) and p2 (bottom) at $t_{MHD}$=0.}
\end{figure}

Figure~\ref{fig:diag_part_t00} presents those quantities for p1 and p2 from Fig.~\ref{fig:traj10p}(a), i.e. for the reference initial setup, focused on one travel between two edges. In this figure as well as 
the following ones, black vertical lines indicate specific times (coinciding with local extrema) in the evolution and facilitate the 
comparison between quantities. In the same spirit, a black horizontal line in each graph indicates the zero value for all quantities. 
It can easily be observed that the curves present similar evolutions with the kinetic energy reaching an extremum at the time the magnitude 
of the electromagnetic fields are maximum. The main difference is the larger value of kinetic energy for p2 than for p1, 
allowing the particle to pass through the equatorial plane instead of undergoing the mirror turn-around like p1. The $x$-component of the electric field seen
by p2 is also 4 times larger, which is in accord with the larger $\mathbf{E}\times\mathbf{B}$ drift visible for its trajectory. Hence, in the initial $t_{MHD}=0$ configuration, we find textbook examples of particle motions, that are merely mirroring due to the $B_z(z)$ variation, augmented with the $\mathbf{E}\times\mathbf{B}$ drift that is essentially along the magnetopause boundary (along $y$). The resistive parallel electric field ($\textit{resist}$) effect is absent (or at round-off level), in accord with collisionless plasma behavior.

\begin{figure}
\centerline{\includegraphics[width=\textwidth]{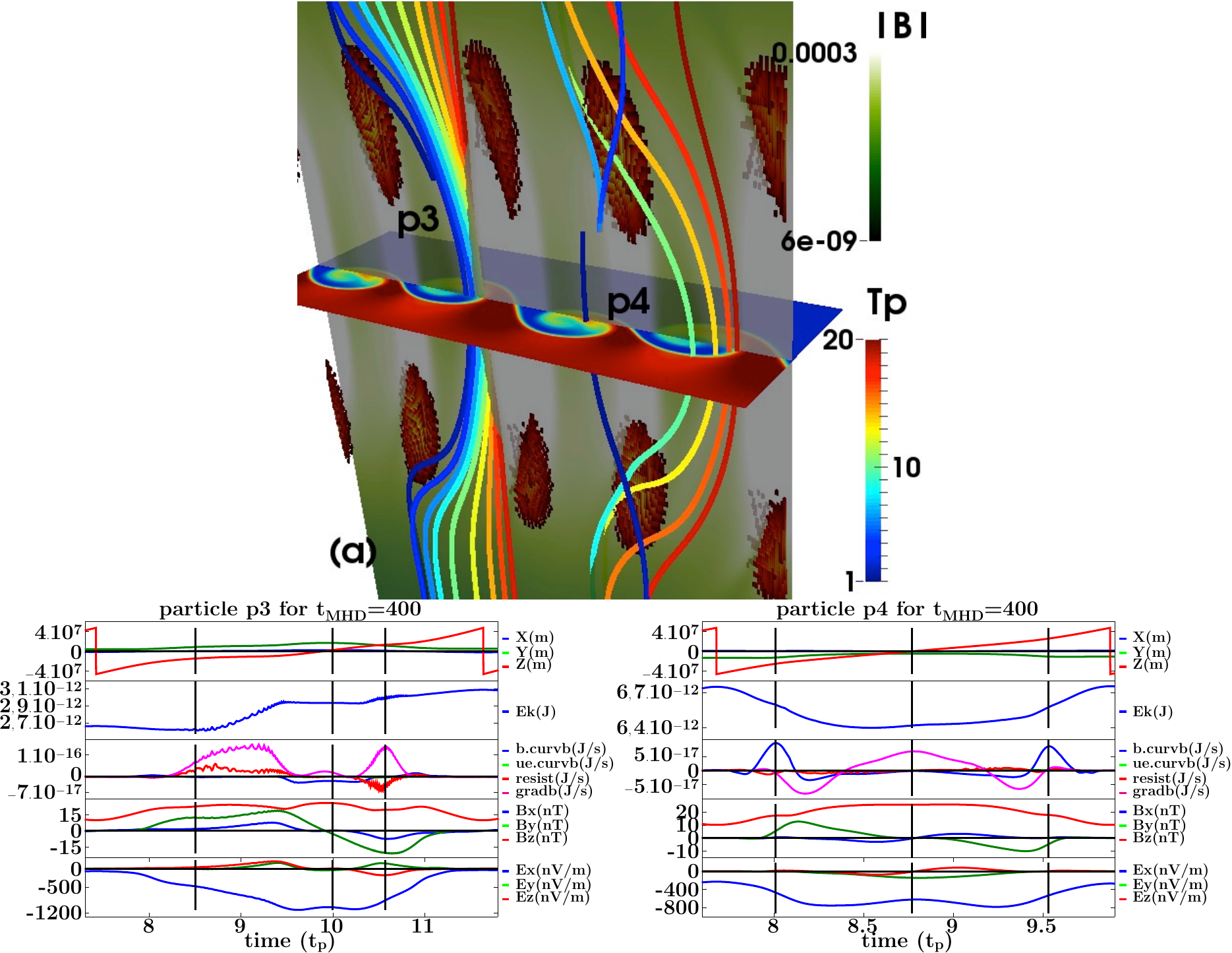}}
\caption{\label{fig:diag_part_t20a}Trajectories (top, coloured with time), positions, kinetic energy, $E_k$ contributions and electromagnetic fields encountered for particles p3 (bottom left) and p4 (bottom right) at $t_{MHD}$=400.}
\end{figure}

In the $t_{MHD}$=400 configuration, particles are exhibiting different mirroring and passing trajectories, as shown for two other particles p3 and p4 in Fig.~\ref{fig:diag_part_t20a}. Particle p4 exhibits 
a bouncing trajectory but passes through the magnetic trap when its angle of approach is slightly different. It represents how the curvature terms are driving particles 
to follow the field lines in their twisted configuration. The rebound takes place when going towards regions of large magnetic field, surrounded by the current sheets, 
while the particle can cross the equatorial plane when it is oriented towards a weaker magnetic field by bending around the current sheets. Particle p3 revisits this 
point in a different fashion and shows how particles coming from positions fairly spread out at the bottom boundary, are funnelled through a particularly narrow area between current sheets. 
The detailed analysis of parts of these trajectories reveals another difference between them. While both p3 and p4 are crossing the equatorial plane from bottom to top and 
evolve through very similar electromagnetic fields, their kinetic energy evolution is different. The curves for p4 display on either side of $z$=0 symmetrical fields 
and contributions, dominated by the \textit{gradb} and \textit{b.curvb} terms. Although its energy is decreased while going from $z$=$-L_z$ to $z$=0, it is restored 
when going from $z$=0 to $z$=$+L_z$. 

It is interesting to notice that the same profile of the magnetic gradient and curvature terms from Eq.~(\ref{Ekinevol}) can either decrease or 
increase the energy of the particle, depending on the direction in which the particle is travelling through the fields. For example, the energy for p3 increases by 15$\%$ 
in two steps coinciding with peaks of the \textit{gradb} and \textit{resist} terms and larger value of the electric field while the particle is at the latitude of 
the current sheets. The increase is larger when both dominating contributions are both positive, indicating that the $\textit{resist}$ term can influence the energy 
both ways for the passing trajectories, while the $\textit{gradb}$ term keeps the same sign as the $y$-component of the electric field. While both particles 
go through the domain on their typically bent trajectories, the resulting variations of kinetic energy and the related contributions can be very different depending on the 
combination of electromagnetic components they encounter.

\begin{figure}
\centerline{\includegraphics[width=\textwidth]{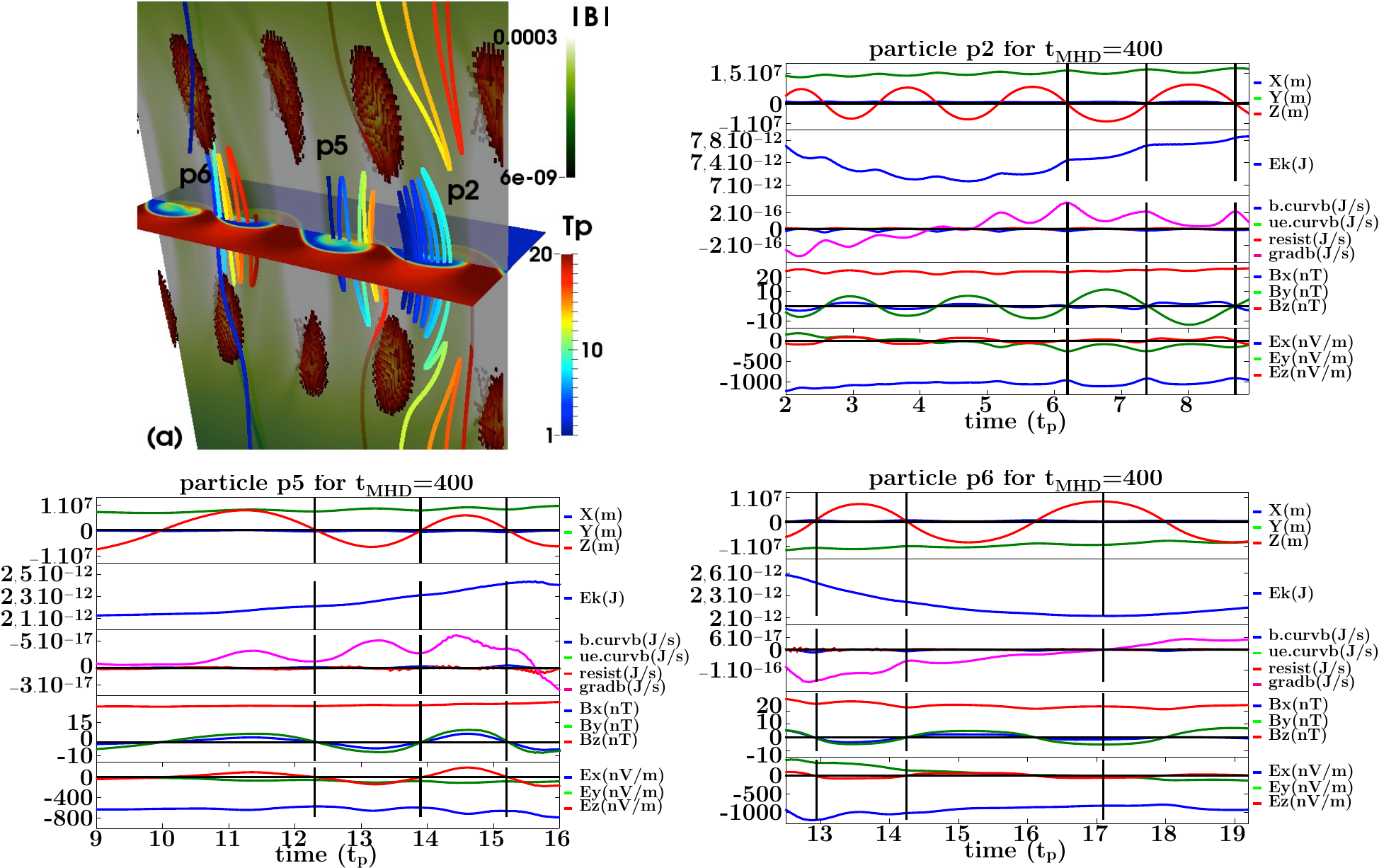}}
\caption{\label{fig:diag_part_t20b}Trajectories (top left, coloured with time), positions, kinetic energy, $E_k$ contributions and electromagnetic fields encountered 
components for particles p2 (top right figure), p5 (bottom left figure) and p6 (bottom right) at $t_{MHD}$=400.}
\end{figure}

The next particles relate to a class of trajectory that is specific for the perturbed MHD configurations. As can be seen in Fig.~\ref{fig:diag_part_t20b}, particles p2 (same as for $t_{MHD}$=0), p5 and p6 are trapped inside 
the equatorial region, either starting there and escaping (p2, p5), or coming from outside and getting trapped (p6). 
Despite sharing this characteristic, these particles still exhibit different behaviours while being trapped. Particle p5 crosses the equatorial plane inside a 
Kelvin-Helmholtz vortex, its trajectory bends to the left and it continuously gains kinetic energy (up to 15$\%$) before escaping next to a current sheet. 
This kinetic energy increase is driven by the positive value of the \textit{gradb} term. Particle 2 is trapped around a low density area, bends to the right and loses 
energy before it gains energy again (13$\%$) when the \textit{gradb} term from Eq.~(\ref{Ekinevol}) becomes positive and finally, it escapes between two current sheets. Lastly, 
particle p6 gets trapped when bouncing near a current sheet to adopt a trajectory similar to p2 and loses kinetic energy as well, before re-gaining energy at late times 
in the simulation. In all cases, the variations of the \textit{gradb} contribution drives the variations of kinetic energy, which increases when this term is positive, 
and decreases when \textit{gradb} is negative. Those variations also appear to be related to the variations of sign of the $y$-component of the electric field.
While the other components, magnetic or electric, are apparently oscillating with no coherence with the \textit{gradb} contribution, this component shows a clear 
link with its variation.

\begin{figure}
\centerline{\includegraphics[width=\textwidth]{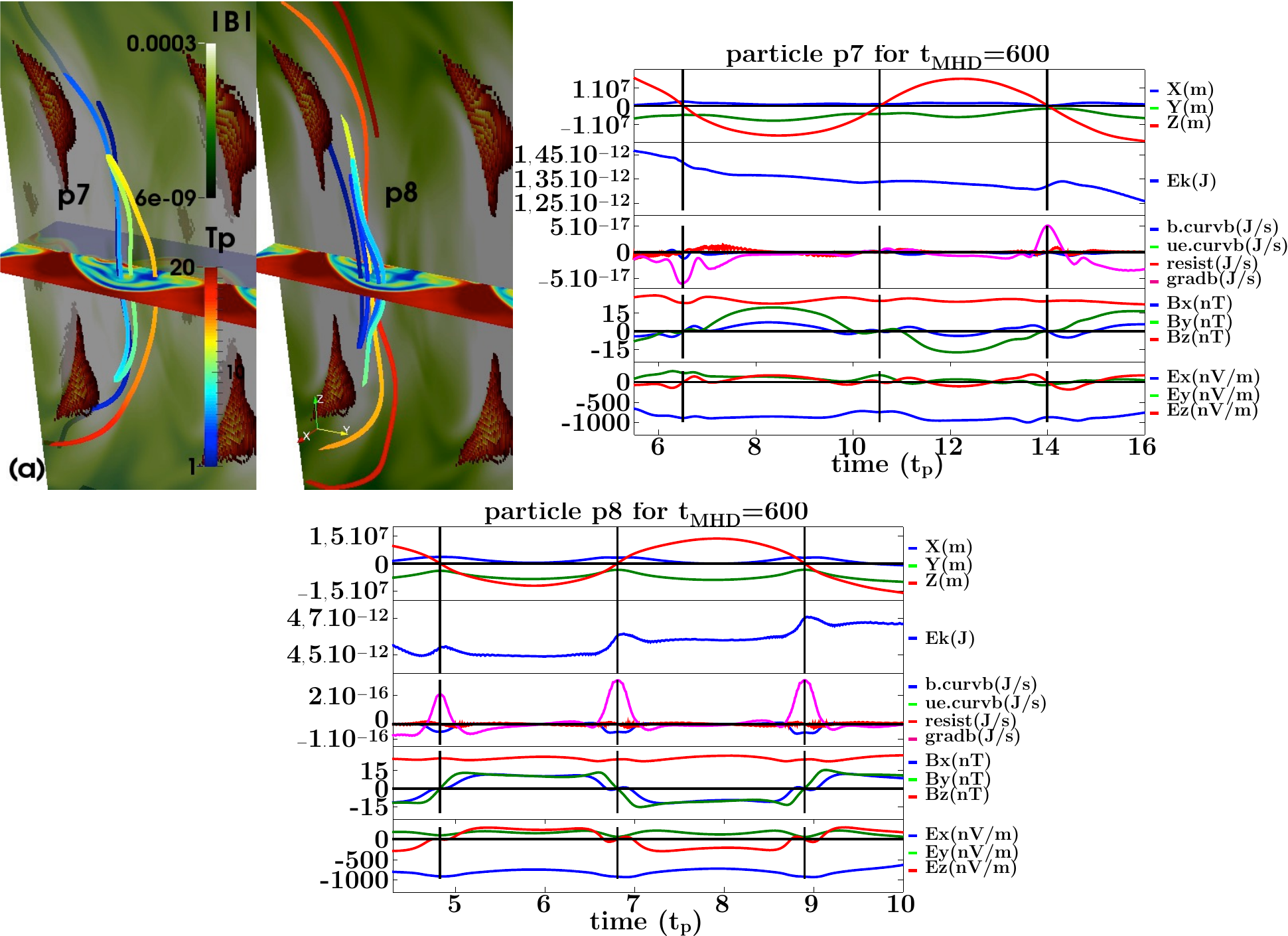}}
\caption{\label{fig:diag_part_t30}Trajectories (top left, coloured with time), positions, kinetic energy, $E_k$ contributions and electromagnetic fields 
encountered for particles p7 (top right) and p8 (bottom) at $t_{MHD}$=600.}
\end{figure}

Finally, Fig.~\ref{fig:diag_part_t30} displays two particles trapped around the equatorial plane for $t_{MHD}$=600. While sharing a very similar trajectory in the same 
region of the MHD background, their kinetic energy variations follow the opposite values of the \textit{gradb} contribution due to the exact topology of the fields 
they cross. Particle p8 demonstrates large energy gain when crossing the $z$=0 plane, due to the large peaks in the \textit{gradb} term, occurring when 
several components of the magnetic and electric fields change sign. Particle p7 sees similar profiles and sign changes but the fact that they do not happen at the 
exact same time (like p8) results in different peaks of the \textit{gradb} term and hence it loses energy each time it crosses the equatorial plane.

\subsection{\label{sssec:catrel}The `catch and release' process}

Up to now, particle trajectories showed only one feature each, either mirroring, passing or in a magnetic trap. A few particles completed trajectories that include a 
full travel through the domain with a trapped episode. Those can give more insight in the process of passing through the equatorial trap and how it affects their 
kinetic energy.

\begin{figure}
\centerline{\includegraphics[width=\textwidth]{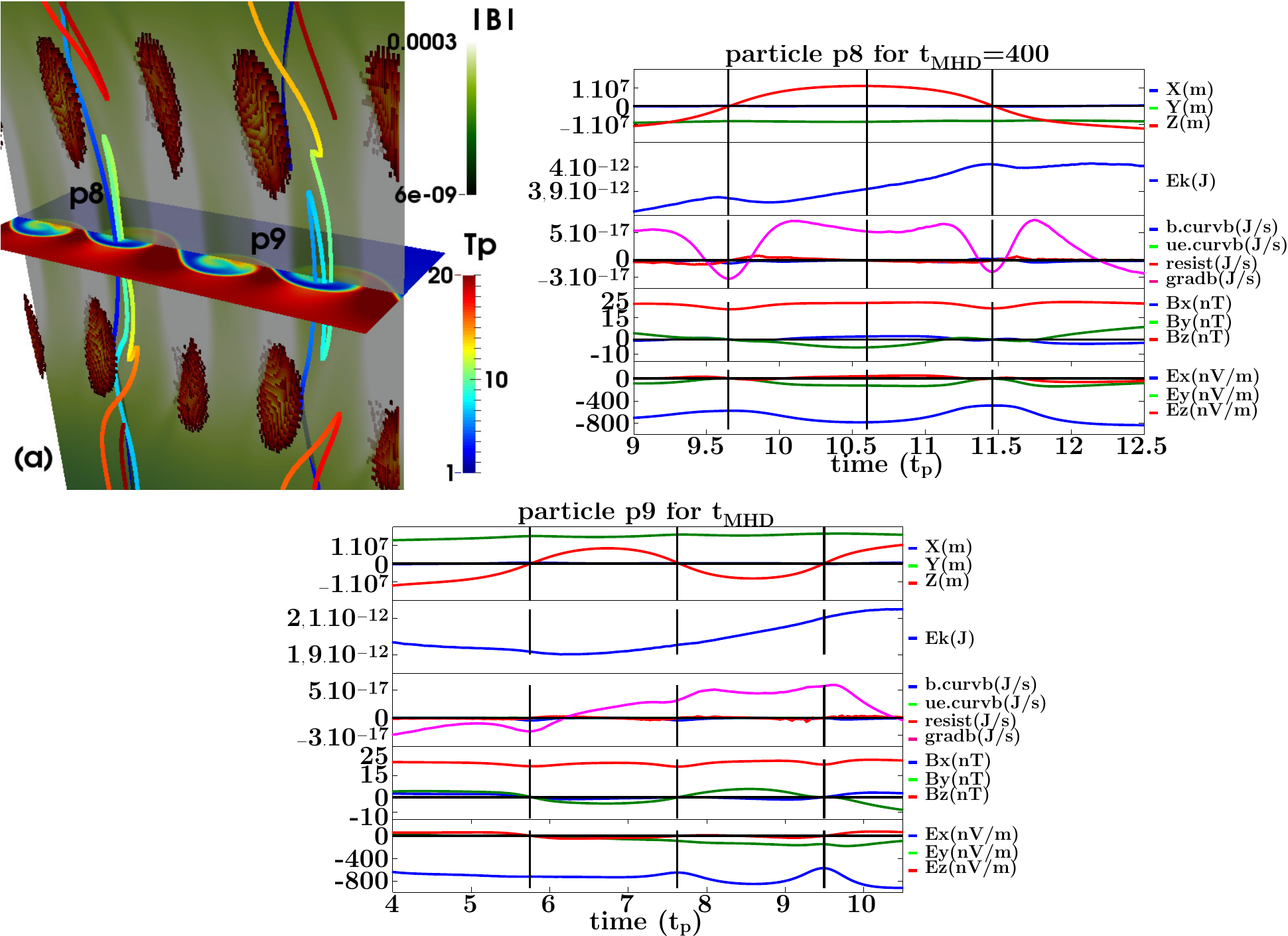}}
\caption{\label{fig:catch_t20}Trajectories (top left, coloured with time), positions, kinetic energy, $E_k$ contributions and electromagnetic fields encountered for particles p8 (top right) and p9 (bottom) at $t_{MHD}$=400.}
\end{figure}

Figure~\ref{fig:catch_t20} displays two particles for $t_{MHD}$=400 that are first getting trapped in the equatorial region, from which they escape afterwards. 
Both have a similar trajectory, entering from the bottom of the region after bending around a current sheet, except for the fact that p9 escapes from the top while 
p8 continues its trajectory returning to the bottom of the domain. Both particles are remaining between the mid-latitudes for a short period, but both exhibit a net 
gain in kinetic energy. This is again correlated with a positive value of the \textit{gradb} term and negative value of the $y$-component of the electric field. 
The same way, p2, p5 and p6 in Fig.~\ref{fig:diag_part_t20b} all showed an energy gain if they remain long enough in the equatorial magnetic trap, even if they 
lost some of it when entering it. It would seem that particles will mostly get accelerated after being caught and released from the equatorial region.

\begin{figure}
\centerline{\includegraphics[width=\textwidth]{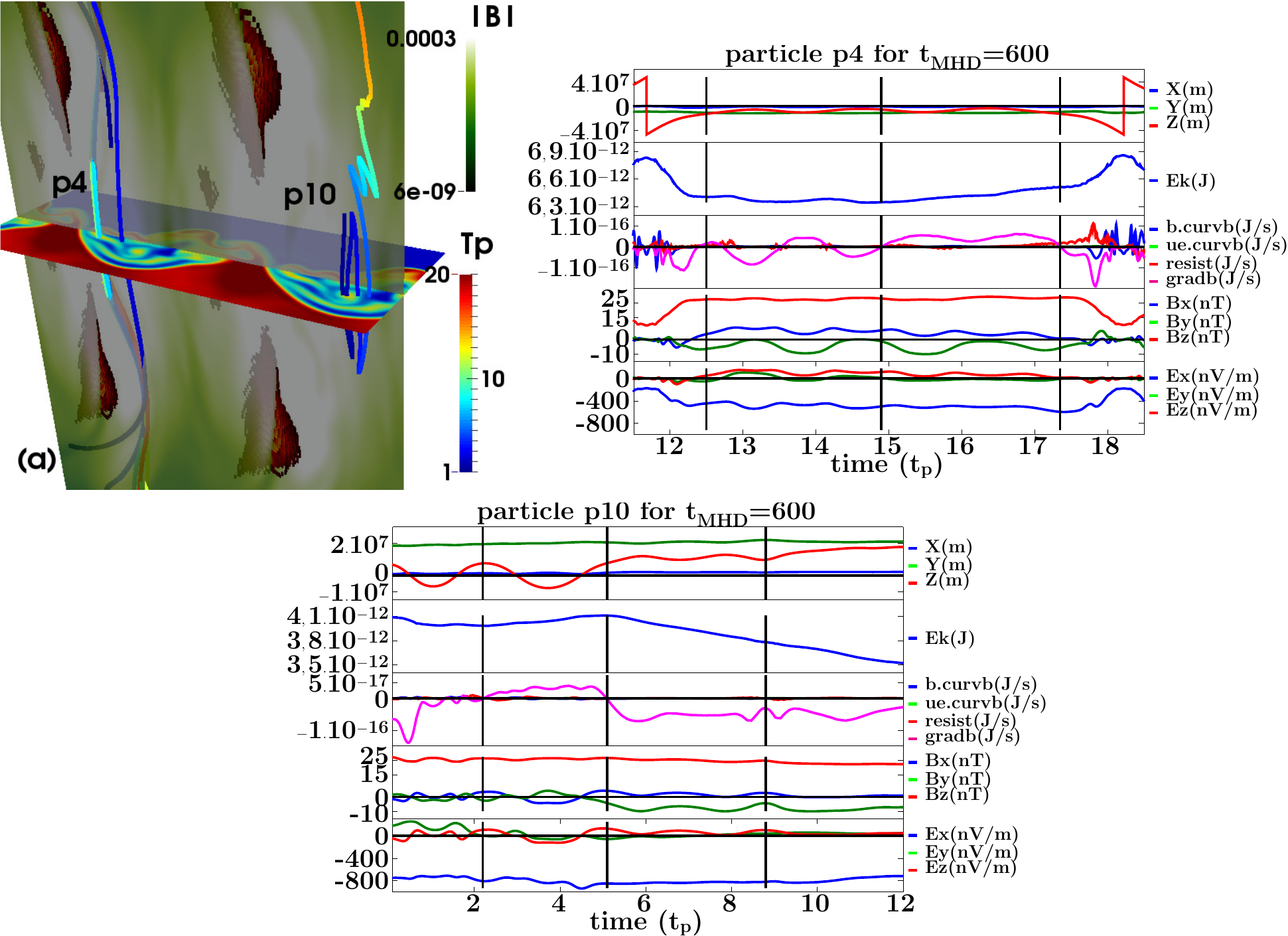}}
\caption{\label{fig:catch_t30}Trajectories (top left, coloured with time), positions, kinetic energy, $E_k$ contributions and electromagnetic fields encountered 
for particles p4 (top right) and p10 (bottom) at $t_{MHD}$=600.}
\end{figure}

This result is confirmed by the particles p4 and p10 in Fig.~\ref{fig:catch_t30} for the nonlinear late KHI stage for $t_{MHD}$=600, where they both exhibit an energy gain while trapped around the equatorial plane. However, 
both particles lose energy when they are caught in the subsequent secondary traps near the current sheets. The variations 
of the kinetic energy are still closely related to the sign of \textit{gradb} and the $y$-component of the electric field. They demonstrate how the catch and release of the particle does not 
always lead to a gain of energy (same as for $t_{MHD}$=0).

\begin{figure}
\centerline{\includegraphics[width=0.8\textwidth]{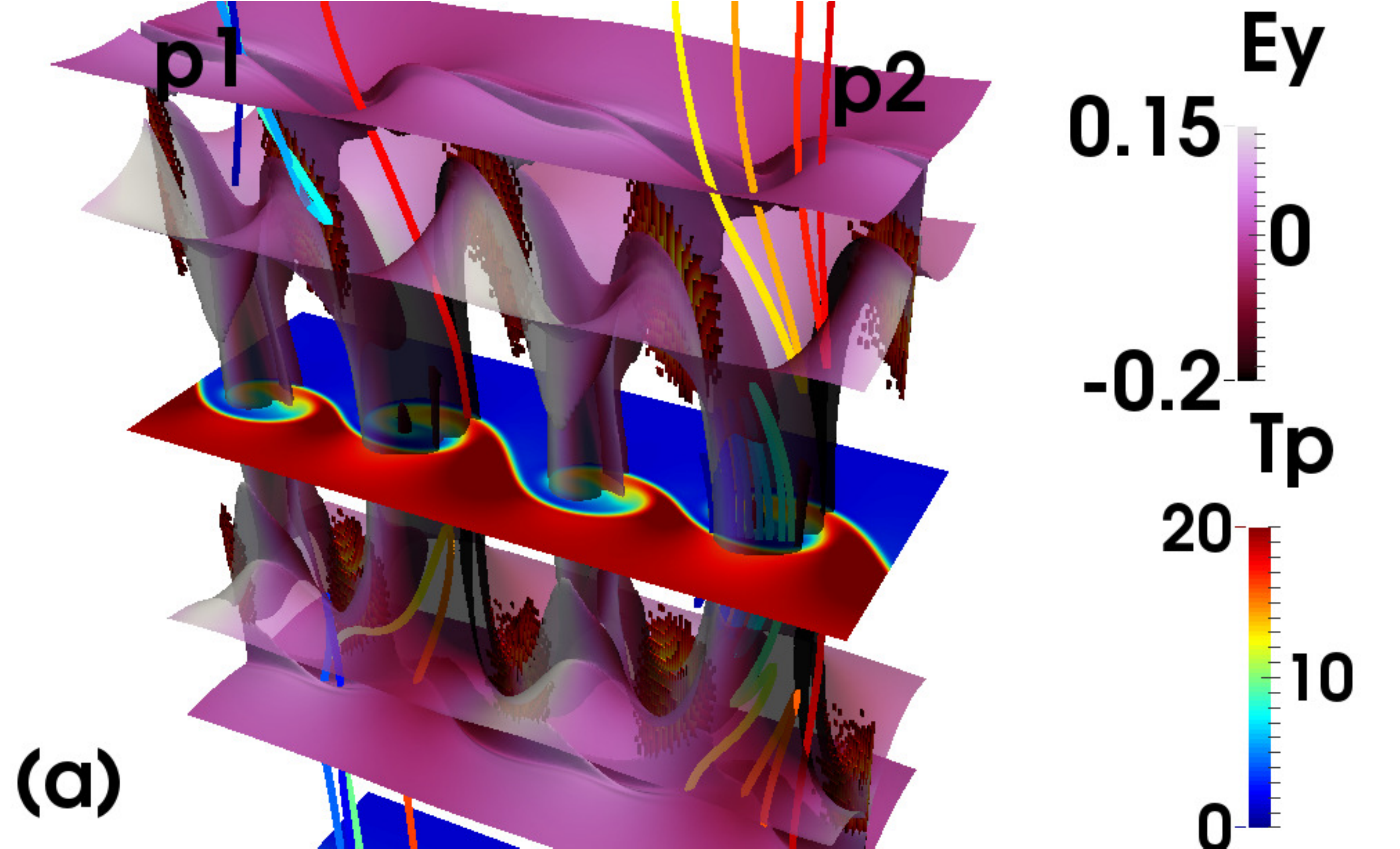}}
\caption{\label{fig:topomag} Magnetic field topology for the DMLR at $t_{MHD}$=400. Two isosurfaces for the magnitude of the magnetic field are represented 
in purple : $|\mathbf{B}|$=0.6 for the external surfaces, $|\mathbf{B}|$=0.8 for the internal surfaces in column shapes. The isosurfaces are coloured with the 
$y$-component of the electric field.}
\end{figure}

Part of the explanation for these different trajectories lies in the perturbations to the configuration of the magnetic field driven by the KHI.
Figure~\ref{fig:topomag} presents an isosurface of the magnetic field, coloured by its $x$-component. The current sheets and passing trajectory particles are included 
as well, to illustrate how particles get trapped or not. By compressing more or less the field lines, the KHI vortices create areas of lower magnetic field, 
connected to narrower areas around the current sheets. Those paths are ways for the particles to pass through the trap, but due to drifting effects, some do not
follow symmetric trajectories. Those particles get caught and they drift until they can escape the trap, most of the time gaining energy in the process.

\section{\label{sec:conclu}Conclusions}

In this paper, electron particle trajectories in the DMLR, as caused by the Kelvin-Helmholtz instability at the magnetopause interface were presented. 
A detailed individual analysis of a series of representative particle trajectories was conducted in order to identify specific acceleration mechanisms. The perpendicular 
magnetic gradient term from the kinetic energy evolution Eq.~(\ref{Ekinevol}) was important for the variations in kinetic energy, with a link to the sign of the $y$-component (along the magnetopause) of the electric field,
though exceptions do exist. In all cases, the parallel electric field acceleration remains small to negligible, in accord with collisionless conditions. Particles can get caught in the areas between the current sheet latitudes during a 
bouncing trajectory, before their release into the outer domain. It appears that the cavities created by the Kelvin-Helmholtz vortices in the equatorial plane can be 
regions where electrons can gain energy. The various sites for trapping particles are clearly related to the current enhancements and the field deformations set up by the KHI development. Both near-equatorial trapping sites, as well as sites at mid-latitude corresponding to the current sheet locations, were identified. From the limited sample of particles that did not leave our domain through the open boundaries at front and back sides of our local box, about 6-10 \% showed trajectories similar to the eight representative particles discussed. We hence estimate that a similar fraction of particles that travel on usual mirror trajectories from pole-to-pole do suffer partial trapping episodes, and notable deflections, during their passage through KHI induced rolls.

This study started from a resistive single fluid MHD simulation, and thus the global dynamics is quantifying 
the ion bulk motion. In our previous work, the effects of the Hall term were studied but it led to minor differences in terms of the KHI development. Particle motions 
were quantified here only within fixed snapshots of the MHD fields, and we can in future work revisit the findings in temporally evolving situations, where magnetic fields can do work. In future work, similar particle orbit studies may be conducted in global 3D simulations, where a more statistical analysis will not be influenced by the open boundary treatments in the local 3D box. An example of such global test particle studies in time-dependent MHD fields is found in~\citeA{Kress2007}, where the observed appearance of a new radiation belt population of very energetic electrons was found consistent with the test particle results. Since it remains challenging to resolve details of the KHI development in global MHD models, our intermediate step for identifying temporary trapping sites in local simulations can aid future analysis of any related particle precipitation. Studying particle orbits is an intermediate step between fluid treatments on the one hand, and more hybrid or fully kinetic treatments on the other hand. The latter do self-consistently treat how fields and particles collectively behave. 

\acknowledgments
Simulations were performed on KU Leuven Tier-1 High Performance Computing cluster Breniac, 
and computational resources and services were provided by the VSC (Flemish Supercomputer Center), funded by the Research Foundation Flanders (FWO) and the 
Flemish Government - department EWI. RK thanks Nanjing University and Purple Mountain Observatory for the kind hospitality during his sabbatical stay. All data were obtained with the open-source code MPI-AMRVAC, which is fully documented and publicly available at {\it{http://amrvac.org}}. BR is supported by an Alexander von Humboldt Fellowship.


%
%


%
%
%
%
%

\end{document}